\documentclass[a4paper,12pt,reqno]{amsart}
\usepackage{mathtools}
\usepackage{etex}
\usepackage{graphicx}
\usepackage{setspace}
\usepackage{amsmath}
\usepackage{amsfonts}
\usepackage{amssymb}
\usepackage{pictex}
\usepackage{amsthm}
\usepackage{graphics,epsfig,verbatim,bm,latexsym,url,amsbsy}
\usepackage{rotating}
\usepackage[authoryear,round]{natbib}
\usepackage{float}
\usepackage{subfig}
\usepackage{mathrsfs}
\usepackage{multirow}
\usepackage{array}
\usepackage{url}
\usepackage[osf]{mathpazo} 
\usepackage[scaled]{helvet} 
\usepackage{hyperref}

\usepackage{xcolor}
\definecolor{myblue}{rgb}{0, 0, 0.54}
\usepackage{hyperref}

\hypersetup{colorlinks=true, urlcolor=myblue, citecolor=myblue, linkcolor=myblue,pdfborder={0 0 0},hyperfootnotes=false}

\usepackage{float}
\usepackage[capitalise,nameinlink]{cleveref}

\crefformat{equation}{#2(#1)#3}
\crefformat{section}{\S#2#1#3}
\crefformat{subsection}{\S#2#1#3}
\crefformat{subsubsection}{\S#2#1#3}
\crefrangeformat{section}{\S\S#3#1#4 to~#5#2#6}
\crefmultiformat{section}{\S\S#2#1#3}{ and~#2#1#3}{, #2#1#3}{ and~#2#1#3}
\crefformat{figure}{#2Figure #1#3}

\usepackage{tikz}
\usepackage{tkz-graph}
\usetikzlibrary{arrows}
\usetikzlibrary{shapes}
\usetikzlibrary{shapes.geometric}

\theoremstyle{definition} \newtheorem{innercustomprop}{Proposition}
\newenvironment{customprop}[1]
{\renewcommand\theinnercustomprop{#1}\innercustomprop}
{\endinnercustomprop}

\theoremstyle{definition} \newtheorem{example}{Example}
\theoremstyle{definition} 
\theoremstyle{definition} \newtheorem{corollary}{Corollary}

\theoremstyle{definition} 
\theoremstyle{definition} \newtheorem{definition}{Definition}
\theoremstyle{definition} 
\theoremstyle{definition} 
\theoremstyle{definition} 
\theoremstyle{definition} 
\theoremstyle{definition} 
\theoremstyle{definition}\newtheorem{proposition}{Proposition}
\theoremstyle{definition} 
\theoremstyle{definition} 
\theoremstyle{definition} 
\theoremstyle{definition} 
\theoremstyle{definition} 
\theoremstyle{definition} 
\theoremstyle{definition} \newtheorem{exercise}{Exercise}

\theoremstyle{definition} \newtheorem{remark}{Remark}

\long\def\symbolfootnote[#1]#2{\begingroup%
\def\thefootnote{\fnsymbol{footnote}}\footnote[#1]{#2}\endgroup}

\usepackage{footnote}
\makesavenoteenv{tabular}
\usepackage{fancyhdr}
\newcommand{\documenttitle}{Thesis}

\newcommand{\be}{\begin{equation}}
\newcommand{\ee}{\end{equation}}
\newcommand{\bes}{\begin{equation*}}
\newcommand{\ees}{\end{equation*}}

\newcommand{\rdots}{\mathinner{%
  \mkern1mu\raise1pt\hbox{.}%
  \mkern2mu\raise4pt\hbox{.}%
  \mkern2mu\raise7pt\vbox{\kern7pt\hbox{.}}\mkern1mu}}

\allowdisplaybreaks

\begin{document}

\title[]{Notes on a Social Transmission Model \\ with a Continuum of Agents}
\author{Benjamin Golub}

\thanks{Department of Economics, Harvard University, bgolub@fas.harvard.edu. I thank Yixi Jiang for exceptional research assistance and Krishna Dasaratha for many helpful conversations. Comments from members of Econ 2034 (Spring 2020) greatly improved the draft.}

\date{\today}

	\maketitle
	
	\section{Introduction}
	
This note presents a simple overlapping-generations (OLG) model of the transmission of a state, such as a behavior,  disease, or awareness of a piece of information. Initially, some fraction of agents carry the trait. In each time period, young agents are ``born''  and are influenced by some older agents. Agents adopt the trait only if at least a certain number of their influencers have the trait. This influence may occur due to rational choice (e.g., because the young agents are playing a coordination game with old agents who are already committed to a strategy), or for some other reason. In any case, our interest is in how the process of social influence unfolds over time, and whether a trait will persist or die out.

Agents may differ both in how many others they are influenced by (their in-degrees), as well as how likely they are to be observed by others (their out-degrees). Our model puts the focus on the heterogeneity in these ``sociability'' attributes, and asks how they affect the long-run fate of the trait in question. Even with a simple model of the network that focuses only on amounts of interaction, the answers are subtle. For example, suppose we perform a mean-preserving spread of influence, making some high-influence agents more influential while low-influence agents become less influential, while the total number of interactions remains fixed. What effect does this have on a trait's likelihood of persisting?

We study the dynamics of transmission and its steady states. Some sharp contrasts can be drawn between two kinds of of contagion. One kind is a \emph{simple contagion}, where being influenced by one person suffices to transmit the trait. Another kind is \emph{complex contagion}, where an agent can only be activated by encountering \emph{multiple} carriers of the trait. While both kinds of contagion can be nested within the same analytical framework, these two types of processes are extremely different in their behavior. Simple contagions can persist starting from a very small population of initial carriers, while complex contagions have a tipping point: they require a critical mass before they are viable. Complex contagions are also more sensitive to the details of interaction: their viability can collapse discontinuously as we increase immunity very slightly. Simple contagions are not susceptible to this sort of ``fragility.''

We derive these results by studying laws of motion that characterize the prevalence of the trait over time, and the steady state. Indeed, if we choose a convenient measure $x_t$ of prevalence, we can describe its evolution  by $$x_t = f(x_{t-1}),$$ where $f$ is a function whose shape, and in particular fixed points, are amenable to simple analysis. This allows for a description of the dynamics of prevalence that is both analytically simple and easy to visualize. The key is to find the right measure of prevalence ($x_t$), and the right $f$, to make this true. This note explains how this is done. This yields a simple and potentially versatile analytical tool.

The main contribution of the model is that, by studying a suitably defined continuum population, tractability can be obtained without any approximation. Many standard models of diffusion, e.g. as surveyed in \citet[Section 7.2]{jackson2008social}, use approximate calculations in large finite networks. The core idea is that branching process ideas help in thinking about large random graphs. But one must then do a fair amount of work to relate heuristic calculations to the behavior of the actual finite-population model that is being studied.\footnote{To our knowledge this has been carried out only for simple contagion in some standard random graph models, but most of the physics literature relies on numerical simulations to validate a mean-field approach.} In the present model, which has a continuum of agents, no approximations are needed, and we can make the analogy between large random graphs and branching processes very tight.

Another advantage of the model, at least from a pedagogical perspective, concerns the way in which sampling biases are handled. Standard expositions of contagion in networks often start with the case of undirected networks, where all contacts are bi-directional. In such models, an agent's opportunities to be influenced are identical to her opportunities to influence others---both occur via her links in an undirected graph. That approach requires a certain subtlety to be dealt with from the very beginning: agents who are exposed to more influence are necessarily disproportionately influential. This ``friendship paradox'' effect is important but creates an additional hurdle for the student. In our exposition, we can start with a simple model where there is no necessary coupling between the propensity to influence and to be influenced. After introducing that simpler case and getting comfortable with the mechanics of the model, we can then move on to the subtleties of the friendship paradox. We can also easily study some alternative assumptions which may be realistic, e.g. that agents who had very many opportunities to be infected may in fact be avoided by others and so less likely to influence them. 

	
	\section{Homogeneous influence}

There is a sequence of cohorts, $N_0, N_1, N_2, \ldots$. For each $t$, the cohort $N_t$ is a copy of the continuum $[0,1]$; its members, called $\emph{agents}$, are labeled $i_t$, where $i \in [0,1]$ and $t$ is the index of the time period. The time-$t$ cohort $N_t$ lives for two periods: at time $t$, its agents are young; they are influenced by elders (members of $N_{t-1}$), and their own state is determined. Then, at time $t+1$, they are old, and their state affects some of the young of the next cohort.

The state in this simple model is binary: some agents are \emph{active} (interpreted as infected, actively manifesting a culture, aware of information, etc.) and others are not. Formally, there is a random variable $A(i_t) \in \{0,1\}$ associated with each agent $i_t$, reflecting whether that agent is active or not. As an initial condition, a fraction $q_0$ of the initial cohort is active.\footnote{We don't care too much which ones. For concreteness, we can say that all $i_0$ with $i\in[0,q_0]$ are active.} 
	
		We begin with a homogeneous version of the model, in which the young sample uniformly from the old.  In other words, old agents do not differ systematically from each other in their propensity to be observed by younger agents. 
	
	For each $t \geq 1$, the timing is as follows: \begin{enumerate}

			\item  For each $i_t \in N_t$,  a set of edges is created.
			\begin{enumerate}
				\item First we randomly draw an in-degree $d_{\text{in}}(i_t)$ for the agent $i_t$, which is distributed according to a probability distribution function $P$ with support on the nonnegative integers.\footnote{$P(d)$ is the probability of having in-degree $d$.} 
			\item We sample  $d_{\text{in}}(i_t)$ agents from the $t-1$ cohort $N_{t-1}$, \textbf{uniformly at random}. For each such agent $j_{t-1}$ sampled, we create a directed influence edge $(j_{t-1},i_t)$. The agents thus sampled are called $i_t$'s \emph{influencers}.
		\end{enumerate}
	The random draws just discussed---the in-degree draws and each agent's sampling of influencers---are independent of each other.\footnote{The independence holds both across different $i_t$ and within a given agent's sampling. There are some technical subtleties having to do with a continuum of random variables, but none that cause any problems for what follows.}
	\item If $A(j_{t-1})=1$ for at least $\tau$ distinct influencers of $i_t$, then $A(i_{t-1})=1$.
	\end{enumerate}

	The evolution of the fraction of actives is the key endogenous variable. Let $q_t$ denote the fraction of agents active at time $t$, or equivalently the probability that an agent sampled uniformly at random is active at time $t$. 
	
	The remaining subsections analyze this model.
	
	\subsection{A simple case: Binomial influence} It is useful to start by considering the case where $P$ is the binomial distribution with $k$ trials and success probability $p$. Here $k$ is a positive integer and $p\in[0,1]$. This case can be interpreted as follows. For each $t \geq 1$, each agent $i_t$ samples $k$ \emph{potential influencers} (uniformly at random from the population, and independently of all others' sampling), and each potential influencer becomes an influencer of $i_t$ with probability $p$, independently. 
	
The special case we have described is called \emph{the $(k,p)$ binomial influence process}. It is useful because it gives a simple one-parameter way to vary $P$ (by varying $p$). We will analyze the evolution of $q_t$ for any given $q_0$ and see how this evolution, and especially the long-run outcome, depends on $p$. Throughout the section, we fix $k$ and treat $p$ as the main parameter.
		
	\begin{example} \label{ex:simple_binomial}
	We begin with the case $\tau=1$. For  $t\geq 1$, \begin{equation} q_t =  1-(1-pq_{t-1})^k. \label{eq:binom_tau_1} \end{equation} The reason is as follows. The agent $i_t$ is active if this agent has at least one potential influencer who becomes an actual influencer and who is active. This combination of events happens for a given potential influencer with probability $pq_{t-1}$. (The first factor is the probability of the potential influencer becoming an actual influencer, and the second is the probability that this member of $N_{t-1}$, sampled uniformly at random, is active.) The quantity $(1-pq_{t-1})^k$ is the probability that the combination fails to happen for each of the $k$ potential influencers. 
	\end{example}

\begin{remark}[No aggregate uncertainty] Note that the evolution of $q_t$ is deterministic. Though individual agents have random outcomes---in terms of whom they observe, whether they become active, etc.---a continuum population ensures that laws of large numbers apply exactly and so the realized fraction of active agents is nonrandom.\end{remark}

\begin{figure}
{	\centering
	\includegraphics[width=0.8\linewidth]{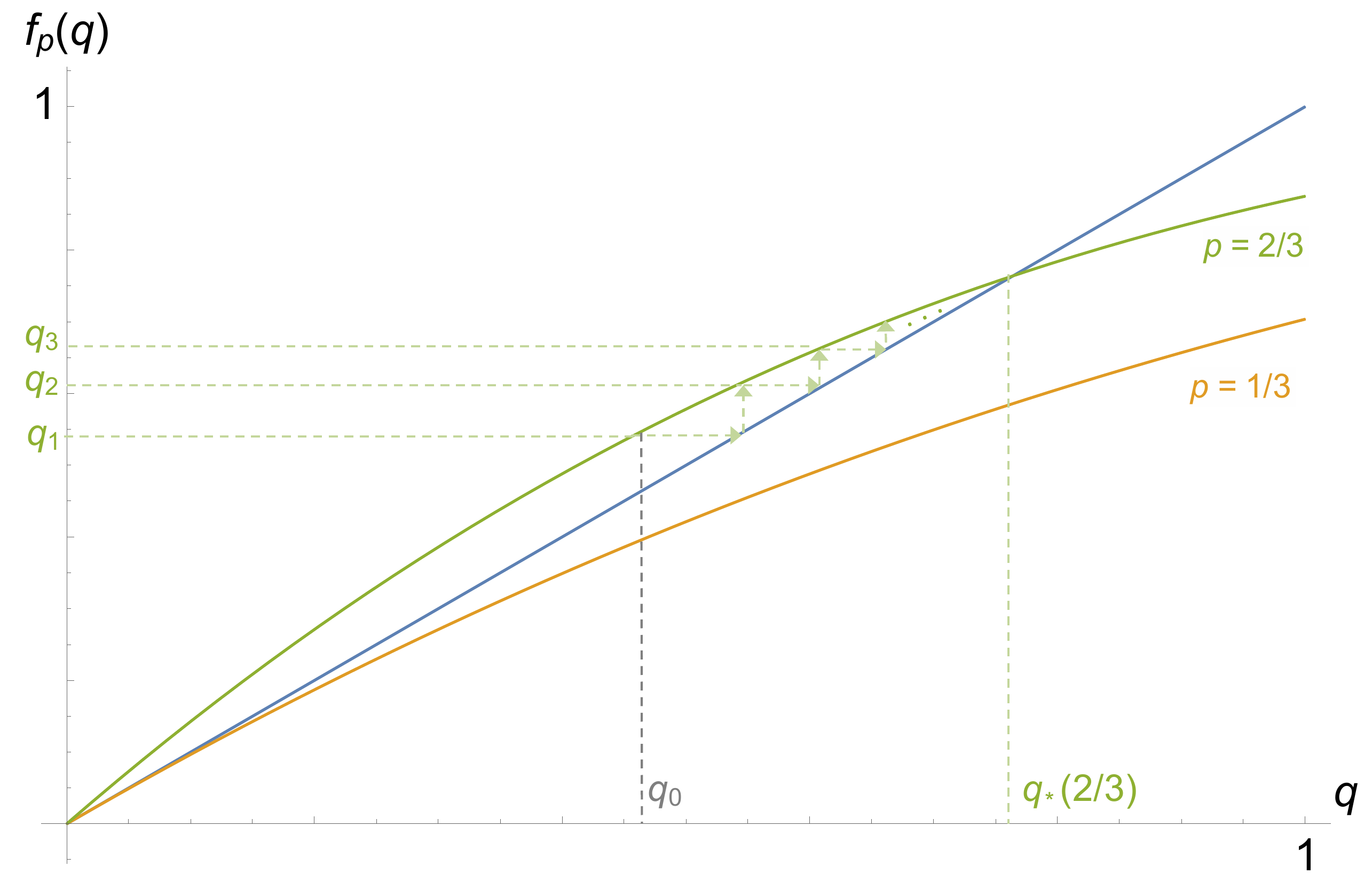}
	\caption{The function $f_p(q)$ in the $\tau=1, k=3$ case for $p=2/3$ (in green) and $p=1/3$ (in orange). The ``staircase" illustrates how we can visualize the sequence defined by $q_t = f_p(q_{t-1})$ from \cref{eq:qt_recursion}, starting from a given $q_0$. Note that the analogous process on the orange curve would converge to $0$. 	\label{fig:simple_contagion_fpq}}}
\end{figure}

By generalizing the logic of \cref{ex:simple_binomial} we deduce:
	\begin{proposition}\label{prop:qt_binom}  Define the function $f_{p,\tau}:[0,1] \to [0,1]$ by \begin{equation}  f_{p,\tau}(q) = \sum_{k'=\tau}^k \binom{k}{k'} (p q)^{k'} (1-p q)^{k-k'}. \label{eq:f_binom} \end{equation} Under the $(k,p)$ binomial influence process with threshold $\tau$, for $t \geq 1$, the fraction $q_t$ of active agents satisfies:
	\begin{equation} \label{eq:qt_recursion} q_t = f_{p,\tau}(q_{t-1}).  \end{equation}
\end{proposition}

We sometimes drop the $\tau$ in the subscript when it is clear from context. In 	\cref{fig:simple_contagion_fpq}, we fix $\tau=1$ and draw two examples of the function $f_{p}$; we also one example of using such a plot to visualize the iteration $q_t = f_{p}(q_{t-1})$ starting from a given $q_0$.

Here are two exercises to help with understanding this basic proposition.

\begin{exercise}
	Show that the dynamic given by equation \cref{eq:binom_tau_1} is a special case of the result in \cref{prop:qt_binom}.
\end{exercise}

\begin{exercise}
	Prove \cref{prop:qt_binom} (at the same level of rigor as our discussion of  \cref{ex:simple_binomial}).
\end{exercise}



Now we turn to analyzing the dynamics of the share of actives.
	
\begin{definition} Let the process start with a fraction $q_0\in [0,1]$ initially infected. Define $$q_\infty(q_0;p)=\lim_{t\to \infty} q_t$$ when the limit exists.  \end{definition}
	
By \cref{prop:qt_binom}, when the limit defining $q_\infty(q_0;p)$ exists, it can be written as  $$q_\infty(q_0;p) = \lim_{t\to \infty} f^t_{p.\tau} (q_0),$$ where $f^t_{p,\tau}$ stands for the function $f_{p,\tau}$ applied $t$ times.

\subsubsection{Dynamics of simple contagion: $\tau=1$}
We now study the case where the threshold is $\tau=1$, so that a single active influencer suffices to activate an agent.	
	
The following proposition gives a characterization of the function $q_\infty(q_0;p)$ in the $\tau=1$ case.
\begin{proposition} \label{prop:tau1_binom_longrun} Let $\tau=1$.
	The quantity $q_\infty(q_0;p)$ is well-defined for all $p\in[0,1]$ and all $q_0\in [0,1]$ and has the following properties: \begin{enumerate}
		\item For all $p\in[0,1]$, we have $q_\infty(0;p)=0$.
		\item For all $p\in[0,1]$, there is a $q_*(p)$ such that $q_\infty(q_0;p)=q_*(p)$ for all $q_0 \in (0,1]$. This $q_*(p)$ is the maximum fixed point of $f_{p}$.\footnote{I.e., the largest $q$ so that $f_{p}(q)=q$.}
	\end{enumerate}
\end{proposition}

In brief, $q=0$ is always a fixed point of the dynamics (though it may be unstable for some values of $p$). If we start from any initial fraction $q_0$ other than $0$, the dynamics converge to $q_*(p)$, the largest fixed point of $f_{p}$, which may be $0$ but, as we will see, is sometimes positive.

\begin{exercise}
	Prove  \cref{prop:tau1_binom_longrun}. 
\end{exercise}

\begin{figure}[H]
	\centering
	\includegraphics[width=0.6\linewidth]{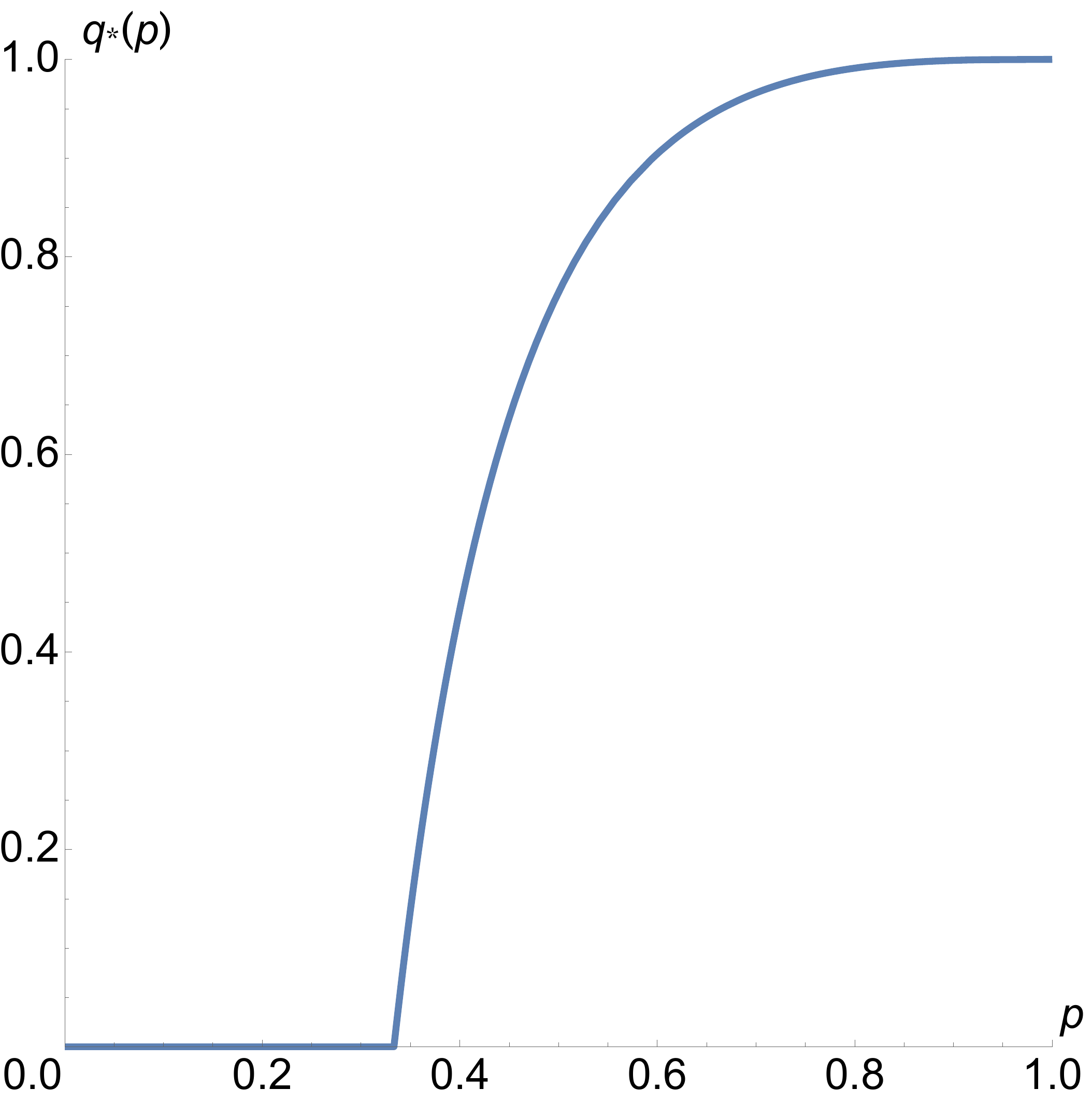}
	\caption{The function $q_*(p)$ in the $\tau=1$ case for $k=3$. }
	\label{fig:simple_contagion_qstarp}
\end{figure}

The next proposition analyzes in more detail this outcome $q_*(p)$.  \cref{fig:simple_contagion_qstarp} depicts the features that the proposition establishes.
\begin{proposition} \label{prop:tau1_binom_qstar} Suppose $\tau=1$. Recall that $q_*(p)$ is the maximum fixed point of $f_p$. Define $\underline{p}=\frac{1}{k}$.  The function  $q_*$ has the following properties: \begin{enumerate}
		\item $q_*$ is a continuous function.
		\item For $p\in[0,\underline{p}]$ we have $q_*(p)=0$.
		\item On the interval $(\underline{p},1]$ the function is strictly increasing, concave, and differentiable.
		\item $\frac{d}{dp} q_*(p) \to \frac{2k^2}{k-1}$ as $p \downarrow \underline{p}$.
	\end{enumerate}
\end{proposition}

The fact that $f_p$ is a concave function for any $p$ ensures that its largest fixed point goes to $0$ continuously as we decrease $p$ to $\underline{p}$.

\begin{exercise}
	Prove \cref{prop:tau1_binom_qstar}. It may help to use the following idea: note that when $$ f_p(q)=1-(1-pq)^k$$ has a strictly positive fixed point $q_*>0$, we can write  $$ q_*=1-(1-pq_*)^k$$ and solve for $p$ as a function of $q_*$. 
\end{exercise}

As a corollary of \cref{prop:qt_binom,,prop:tau1_binom_longrun,,prop:tau1_binom_qstar} we can give a complete description of the dynamics of the $q_t$.
\begin{corollary} The dynamics defined by \cref{eq:qt_recursion} have the following properties:
\begin{enumerate}
	\item Suppose $p \in [0, \underline{p}]$. If $q_0>0$, then $q_t$ converges to $0$ monotonically. Thus $0$ is the unique, globally stable fixed point of the dynamics.
	\item Suppose $p \in (\underline{p},1]$. If $q_0>0$, then $q_t$ converges to $q_*(p)>0$ monotonically. Thus, $q_*(p)$ is the unique stable fixed point of the dynamics, while $0$ is an unstable fixed point. 
\end{enumerate}
\end{corollary}

\subsubsection{Dynamics of complex contagion: $\tau>1$}		We now take a brief look at the case where the threshold is $\tau>1$, so that an agent must have multiple active influencers to become activated. 

\begin{figure}
	\centering
	\includegraphics[width=0.8\linewidth]{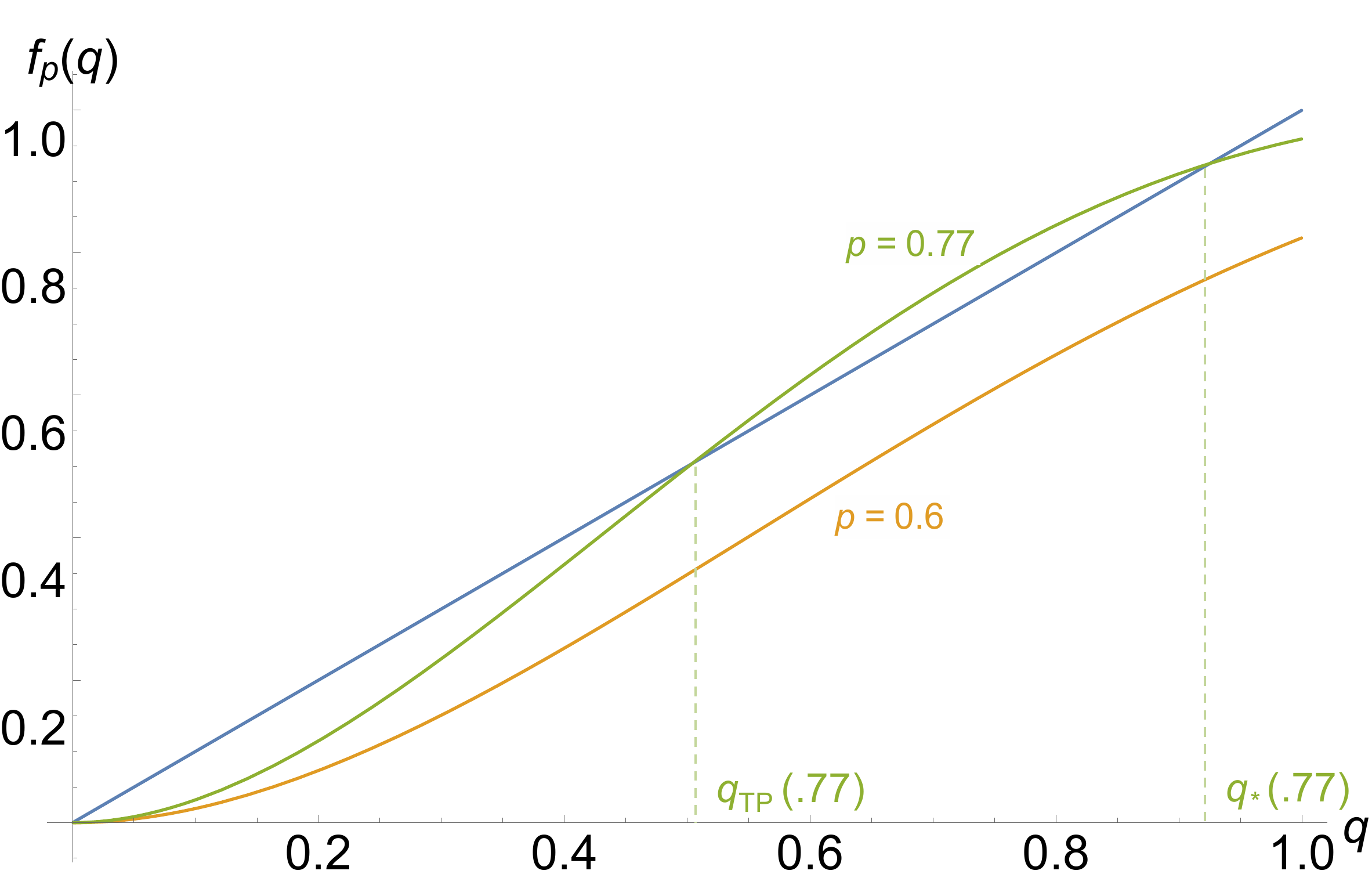}
	\caption{The function $f_{p,\tau}(q)$ when $k=4$ and $\tau=2$. }
	\label{fig:complex_contagion_fpq}
\end{figure}

Because $f_{p,\tau}$ is now $S$-shaped, as depicted in  	\cref{fig:complex_contagion_fpq}, the dynamics are now more complicated. First, we document how the fixed points of $f_{p,\tau}$ depend on $p$, which is the analogue of  \cref{prop:tau1_binom_qstar}.

\begin{proposition} \label{prop:taubig_binom_qstar} Suppose $\tau>1$. There is a value\footnote{We drop the arguments on it in the statements below.} $\underline{p}(k,\tau)>0$ such that \begin{enumerate}
		\item For $p\in[0,\underline{p}]$ the only fixed point of the function $f_{p,\tau}$ is $0$.
		\item There are two differentiable functions $q_1,q_2:[\underline{p},1] \to [0,1]$ such that, for $p\in[\underline{p},1]$, we have $$f_{p,\tau}(q)=q \quad \text{ for } \quad q=0,\,q_1(p),\, q_2(p).$$
		\begin{enumerate}
			\item If $p>\underline{p}$, then 	$9<q_{1}(p) < q_{2}(p)$ and $f_{p,\tau}$ has three distinct fixed points.
			\item If $p=\underline{p}$ then $0<q_{1}(p) = q_{2}(p)$ and $f_{p,\tau}$ has two distinct fixed points.
	 \end{enumerate}		
 	\item $q_1$ is strictly decreasing and $q_2$ is strictly increasing.
 	\item $q_2(\underline{p})>0$ and  $\frac{d}{dp} q_2(p) \to \infty$ as $p \downarrow \underline{p}$.
	\end{enumerate}
\end{proposition}

\begin{exercise}
	Prove  \cref{prop:taubig_binom_qstar}. A suggestion: take for granted that 
$$ f_{p,\tau}''(q)	= \frac{\tau  \Gamma (k+1) (p q)^{\tau } (1-p q)^{k-\tau -1} }{q^2 \Gamma (\tau +1) \Gamma (k-\tau +1)} \left[ \tau +1-(k-1) p q\right],$$ where $\Gamma$ is the Gamma function and deduce from this  that $f_{p,\tau}$ has at most one inflection point.
\end{exercise}

With this result in hand, by thinking about the dynamics of $f_{p,\tau}^t$ for the two types of curves plotted  in  	\cref{fig:complex_contagion_fpq},\footnote{As well as the case where the curve is tangent to the diagonal line.} we can deduce the following.

\begin{proposition} \label{prop:taubig_binom_dynamics} Assume $\tau>1$.  \begin{enumerate}
		\item Suppose $p\in[0,\underline{p}]$. The only fixed point of the function $f_{p,\tau}$ is $0$ and this fixed point is globally stable.
		\item Suppose $p\in[\underline{p},1]$. If $q_0> q_{1}(p)$ then the dynamics converge monotonically to $q_{2}(p)$ and if $q_0 < q_1(p)$ the dynamics converge monotonically to $0$. Thus the basin of attraction of the fixed point $0$ is $[0,q_1(p))$. 
		\begin{enumerate} \item If $p>\underline{p}$, the basin of attraction of the fixed point $q_2(p)$ is $(q_1(p),1]$. The fixed point $q_1(p)$ is unstable. 
		\item  If $p=\underline{p}$, then because $q_1(p)=q_2(p)$, the basin of attraction of the fixed point $q_2(p)$ is $[q_1(p),1]$. The fixed point $q_1(p)=q_2(p)$ is half-stable. \end{enumerate} 
	\end{enumerate}
\end{proposition}

One quick way to summarize this result is that $q_1(p)$ is a \emph{tipping point}: if we start at a $q_0$ below it, then the dynamics converge to $0$, but if we start above it, then the dynamics converge to $q_2(p)$. There was no analogue of this in the $\tau=1$ model; there, any positive $q_0$ led to the positive fixed point of $f_{p,\tau}$ when there was one.

\subsection{Analysis for a general in-degree distribution} \label{sec:in_deg_gen} We will now examine the case of a general $P$. 

	\begin{proposition}\label{prop:qt_gen}  Define the function $f_{P,\tau}:[0,1] \to [0,1]$ by \begin{equation} f_{P,\tau}(q) =  \sum_{d=\tau}^\infty P(d) \sum_{k'=\tau}^d \binom{d}{k'} q^{k'} (1-q)^{d-k'}.  \label{eq:qt_gen} \end{equation} Under the homogeneous sampling model, the fraction $q_t$ of active agents satisfies:
	\begin{equation} \label{eq:qt_iteration_general} q_t = f_{P,\tau}(q_{t-1}).  \end{equation}
\end{proposition}

The proposition characterizes the dynamics of $q_t$ for arbitrary in-degree distributions $P$. We now explain this characterization. Let us focus on an agent with in-degree $d$ and compute $q_{t,d}$, the probability that this individual is activated. This agent's influencers are drawn uniformly at random from $N_{t-1}$, and thus are active with probability $q_{t-1}$. It follows that  \begin{equation} \label{eq:qtd} q_{t,d}= \sum_{k'=\tau}^d \binom{d}{k'} q_{t-1}^{k'} (1-q_{t-1})^{d-k'}.\end{equation} On the right-hand side we have simply written out the probability that a Bernoulli random variable with success probability $q_{t-1}$ and $d$ total trials has at least $\tau$ successful trials; here ``success'' corresponds to an influencer being active. To compute $q_t$, which is the probability that a \emph{randomly-selected} individual is activated, we simply average these according to the degree distribution: $$ q_t = \sum_{d=\tau}^\infty P(d) q_{t,d}. $$

\begin{example} \label{ex:tau1generating} In the special case $\tau=1$, we may write (dropping the $\tau$ argument)
$$ f_P(q) = \sum_{d=0}^\infty P(d)[1-(1-q)^{d}] =    1- \sum_{d=0}^\infty P(d)(1-q)^{d} .  $$ Recalling that the generating function of the distribution $P$ is the series \begin{equation}\mathcal{G}_P(x)=\sum_{d=0}^\infty P(d)x^d, \label{eq:generating_function_definition} \end{equation} we have \begin{equation} f_P(q) =  1- \mathcal{G}_P(1-q). \label{eq:generating_function_tau_1} \end{equation} \end{example}

This example motivates a restatement of  \cref{prop:qt_gen}. It will be helpful to make a definition: 
\begin{definition}[Generalized generating function]
\begin{equation}
{G}_{P,\tau}(x)=\sum_{d=0}^\infty P(d) \sum_{k'=0}^{\tau-1} \binom{d}{k'} (1-x)^{k'} x^{d-k'}. 
\label{eq:generalized_generating}
\end{equation} 
\end{definition}
This is a generalization of the ordinary generating function because $\mathcal{G}_{P,1}=\mathcal{G}_P$ as defined in \cref{eq:generating_function_definition}. Noting that 	equation \cref{eq:qt_gen} can be rewritten as $$ f_{P,\tau}(q)=1-\mathcal{G}_{P,\tau}(1-q), $$ we then have the following restatement of  \cref{prop:qt_gen}.
\begin{customprop}{\ref*{prop:qt_gen}'} \label{eight}
Define the function $f_{P,\tau}:[0,1] \to [0,1]$ by \begin{equation} f_{P,\tau}({q}_t) = 1-\mathcal{G}_{P,\tau}(1-{q}_t).  \label{eq:qt_gen_genfunction} \end{equation} Under the homogeneous sampling model, the fraction ${q}_t$ of active agents satisfies:
$$ \label{eq:qt_iteration_general_genfunction} {q}_t = f_{P,\tau}({q}_{t-1}). $$
\end{customprop}

\subsubsection{Immunity as a parameter} In the $(k,p)$ binomial model, we had a straightforward way of varying the contagiousness of the state: varying $p$. Now there is no direct analogue of $p$. However, we can change the model by stipulating that a fraction $\pi$ of the nodes in each cohort are exogenously immune (i.e, cannot be active) and the rest---a fraction $\overline{\pi}$---are susceptible, behaving exactly as in the basic model.  The immune nodes effectively become nodes with in-degree $0$, and the rest of $P$ is correspondingly scaled down. 

Instead of  \cref{eq:qt_gen_genfunction}, we now have \begin{equation} f_{P,\tau}(q;\overline{\pi}) = \overline{\pi}\left[  1- \mathcal{G}_{P,\tau}(1-q) \right] \label{eq:qt_gen_genfunction_pi}. \end{equation} 
The dynamics are given by $q_{t}=f_{P,\tau}(q_{t-1};\overline{\pi})$. 
Now we can treat $\overline{\pi}$ as a parameter to vary, and carry out exercises similar to those we did above when we varied $p$. 

\begin{exercise} Assuming that $q$ is a positive solution of $q=f_{P,\tau}(q;\overline{\pi})$ in  \cref{eq:qt_gen_genfunction_pi}, write $\overline{\pi}$ as a function of $q$. Use this to plot all fixed points of $f_{P,\tau}(\cdot;\overline{\pi})$ as a function of $\overline{\pi}$ when  $P$ is Binomial$(k,p)$ and $\tau=2$. \end{exercise}

\subsubsection{Analogy with a branching process} Note that for $\tau=1$, the dynamic \cref{eq:qt_iteration_general} is closely related to the classic Galton-Watson branching process, and the active fractions $q_t$ have a simple interpretation in terms of this process. A node $i_t$ has  influencers (analogous to children in the Galton-Watson process) whose number is distributed according to $P$.  These influencers, $j_{t-1}$, have influencers of their own, and so on. Let $T(i_t)$ be the union of all paths into $i_t$ in the (random) influence graph, which is an \emph{arborescence}.\footnote{A directed graph in which every node has exactly one directed path to the root, $i_t$. This is basically a tree rooted at $i_t$, where all edges are directed toward $i_t$.} The agent $i_t$ is active if and only if at least one node in this arborescence is an active agent in $N_0$.  If $q_0=1$, then $q_t$ is simply the probability of the arborescence of indirect influence not dying out before it goes back $t$ generations. It can be seen that this is the probability of a Galton-Watson process, where each node draws a number of children from $P$, surviving for $t$ generations. If $q_0<1$, then $i_t$ being active requires something more stringent---that one of the indirect influencers ``hit'' by the Galton-Watson tree at the ``last'' (i.e., oldest, farthest-back) layer is one of those that was exogenously set to be active.

\section{Heterogeneous influence}

The basic setup is the same in terms of the structure of the overlapping generations model. In the previous section, all agents in $N_{t-1}$ had the same ex ante probability of being sampled by an $i_t \in N_t$. Now, however, different agents will have different probabilities of being sampled, and this will affect the probability of a typical influence edge carrying the contagion.

We first explain what is the key new moving part we must introduce. Continuing for now with the model of the previous section, recall $q_{t-1,d}$ is the probability that a $j_{t-1} \in N_{t-1}$ with in-degree $d$ is active.  Equation  \cref{eq:qtd} states that this number depends on $d$: someone at $N_{t-1}$ who had more influencers is likelier to be active. We did not spend a lot of time keeping track of these numbers separately; we just averaged them (weighted by $P(d)$) and focused on \begin{equation} \label{eq:qt-1simple} q_{t-1}=\sum_d P(d) q_{t-1,d}.\end{equation} This was because everyone in $N_t$ sampled influencers uniformly at random; an influencer's probability of being sampled was independent of her in-degree $d$. Since $j_{t-1}$'s probability of being sampled is proportional to her expected \emph{out}-degree, an equivalent statement of the assumption is that out-degree is uncorrelated with in-degree. In contrast, in this section we will allow an influencer $j_{t-1}$'s probability of being sampled to depend on  \emph{$j_{t-1}$'s own in-degree}, $d_{\text{in}}(j_{t-1})$. That is, we are allowing $j_{t-1}$'s out-degree to be correlated with in-degree. In this case, the $q_{t-1}$ in \cref{eq:qt-1simple} is no longer the probability an influence edge comes from an active agent, as it was in the last section. We must account for the non-uniform sampling; some $q_{t-1,d}$'s may need to be over-weighted, and others under-weighted, because the corresponding agents are systematically over-sampled or under-sampled. This section discusses how to adjust the model and the analysis to account for such effects.

We first formalize the timing of the richer setting. For each $t\geq 1$: \begin{enumerate}

\item  For each $i_t \in N_t$,  a set of edges is created.
	\begin{enumerate}
		\item First we randomly draw in-degree $d_{\text{in}}(i_t)$ for the agent $i_t$, which is distributed according to a probability distribution function $P$ with support on the nonnegative integers.
		\item  We sample  $d_{\text{in}}(i_t)$ agents from the $t-1$ cohort $N_{t-1}$. For each such agent $j_{t-1}$ sampled,  we create a directed influence edge $(j_{t-1},i_t)$. These are called $i_t$'s \emph{influencers}. The probability of $j_t$ being sampled depends  $j_t$'s in-degree. Let $P_{\text{infl}}(d')$ be the probability of an agent $j_{t-1}$ with in-degree $d_{\text{in}}(j_{t-1})=d'$ being sampled by any $i_t$.\footnote{Note that mechanically, the probability of $j_{t-1}$ being sampled is proportional to her expected  \emph{out}-degree (which is a quantity we have not introduced notation for). The distribution $P_{\text{infl}}$ tracks whether this sampling probability is \emph{also} correlated with $j_{t-1}$'s in-degree. Since $j_{t-1}$'s activity is predicted by $j_{t-1}$'s in-degree, not her out-degree, we will see that it is the information contained in $P_{\text{infl}}$ that we ultimately care about. Note that $P_{\text{infl}}$ can be quite different from $P$.} We call $P_{\text{infl}}$ the \emph{influencer in-degree distribution}.
	\end{enumerate}
	The random draws just discussed---the in-degree draws and each agent's sampling of influencers---are independent of each other.\footnote{The independence holds both across different $i_t$ and within a given agent's sampling.}
	\item If $A(j_{t-1})=1$ for at least $\tau$ distinct influencers of $i_t$, then $A(i_{t-1})=1$.
\end{enumerate} 

What is key to this model being as tractable as that of the previous section is that every agent in $N_t$ samples elders, independently, in the same way. However,  some agents in $N_t$ may sample more (i.e., may have a higher in-degree) than others, and as we have emphasized, their in-degrees may now be correlated with their propensity to be sampled by others.

\begin{exercise} Give a precise description of an environment similar to the above with the following properties: \begin{enumerate} \item[(i)] agents' expected \emph{out-degrees} (i.e., number of agents they influence) are different (i.e., there are multiple types of agents, each with a different expected out-degree);
		\item[(ii)] the probability of an influence edge coming from an active agent is the $q_t$ of the previous section. \end{enumerate} You will need to define an extension of the above model rather than a special case. Your example will illustrate why out-degree \emph{per se} does not matter---only its correlation with in-degree. \end{exercise}

\begin{example}[Influence proportional to in-degree] There is a special but important kind of $P_{\text{infl}}$ to consider, because it comes up a lot in random graph theory.  Suppose an agent's expected out-degree is equal to her  in-degree.  In this case, the probability of $j_{t-1}$ with in-degree $d'$ being sampled is proportional to $P(d')$, the fraction of agents who have this degree, and also proportional to $d'$. The latter proportionality holds because if we double $d'$, we double the out-degree, and thus this degree-type's opportunities for influence; it must then be twice as likely to be drawn as an influencer. The distribution $\widetilde{P}$ is defined by $\widetilde{P}(d) \propto d P(d)$, or if we do the normalization explicitly, $$ \widetilde{P}(d) = \frac{d P(d)}{\sum_d dP(d)}.$$ \end{example}

\subsection{Analysis}
Let $q_{t,d}$ be the fraction of agents in $N_t$ with in-degree $d$ who are active. Define \begin{equation} \label{eq:qthat_def} \widehat{q}_{t} =  \sum_{d'} P_{\text{infl}}(d') q_{t,d'} \end{equation} to be the  expected activity of an individual sampled from the influencer in-degree distribution, which we call the \emph{influence-weighted activity}. Finally, recall the definition of $f_{P,\tau}:[0,1] \to [0,1]$ from  \cref{sec:in_deg_gen}, e.g., \cref{eq:qt_gen_genfunction}.  
	\begin{proposition}\label{prop:qt_gen_het}   Under the non-homogenous sampling model, we have
	\begin{equation}  \label{eq:qt_dynamics} \widehat{q}_t = f_{P_{\text{infl}},\tau}(\widehat{q}_{t-1}).  \end{equation} Moreover, \begin{equation}\label{eq:qtd_general}q_{t,d}= \sum_{k'=\tau}^d \binom{d}{k'} \widehat{q}_{t-1}^{k'} (1-\widehat{q}_{t-1})^{d-k'}. \end{equation}  Therefore, the sequence $(\widehat{q}_t)_{t=0}^\infty$ whose evolution we have characterized in \cref{eq:qt_dynamics} allows us to compute any  $q_{t,d}$.
\end{proposition}

Note that the proposition focuses on $\widehat{q}_t$ rather than $q_t$. But we can easily compute $q_t$ once we know the $q_{t,d}$, using the formula $q_t = \sum_d P(d) q_{t,d}$.

We now explain why the proposition is true. Consider an agent at time $t$ with in-degree $d$. This agent's influencers are drawn from $N_{t-1}$ and those who themselves had in-degree $d'$ are sampled with probability $P_{\text{infl}}(d')$. It follows that the probability of a random influencer being active is $$\widehat{q}_{t-1} =  \sum_{d'} P_{\text{infl}}(d') q_{t-1,d'}.$$ From this we deduce that $$q_{t,d}= \sum_{k'=\tau}^d \binom{d}{k'} \widehat{q}_{t-1}^{k'} (1-\widehat{q}_{t-1})^{d-k'}.$$ On the right-hand side we have simply written out the probability that a Bernoulli random variable with success probability $\widehat{q}_{t-1}$ and $d$ total trials has at least $\tau$ successful trials; here ``success'' corresponds to an influencer being active. 

Now, in order to characterize the  dynamics, we will take a weighted sum of equations \cref{eq:qtd_general} so that we get a $\widehat{q}_t$ on the left-hand side. Multiplying the $q_{t,d}$ equation by $P_{\text{infl}}(d)$ and adding up all these equations, we get 
\begin{equation}\label{eq:qthat_general}\sum_d P_{\text{infl}}(d)q_{t,d}= \sum_d P_{\text{infl}}(d)\sum_{k'=\tau}^d \binom{d}{k'} \widehat{q}_{t-1}^{k'} (1-\widehat{q}_{t-1})^{d-k'}.\end{equation}
In other words:
\begin{equation}\label{eq:qthat_general2}\widehat{q}_t= \sum_d P_{\text{infl}}(d)\sum_{k'=\tau}^d \binom{d}{k'} \widehat{q}_{t-1}^{k'} (1-\widehat{q}_{t-1})^{d-k'}.\end{equation}

\bibliographystyle{ecta}
\bibliography{culture}
	
\appendix

\end{document}